# REVIEW

# Disorder by Design: Unveiling Local Structure and Functional Insights in High Entropy Oxides

John P. Barber[a], William J. Deary[a], Andrew N. Titus[a], Gerald R. Bejger[a], Saeed S.I. Almishal[b], Christina M. Rost[a, †]

High entropy oxides (HEOs) are a rapidly growing class of compositionally complex ceramics in which configurational disorder is engineered to unlock novel functionality. While average crystallographic symmetry is often retained, local structural and chemical disorder, including cation size and valence mismatch, oxygen sublattice distortions, and site-specific bonding, strongly governs ionic transport, redox behavior, magnetic ordering, and dielectric response. This review outlines how these modes of disorder manifest across key oxide families such as rock salt, spinel, fluorite, and perovskite. We highlight recent advances in spectroscopy, total scattering, and high-resolution microscopy enable multi-scale insight into short- and intermediate-range order. By integrating experimental observations with theoretical modeling of entropy and local energetics, we establish a framework linking structural heterogeneity to emergent properties. These insights not only deepen our fundamental understanding of disorder–property relationships but also offer a path toward rational design of tunable materials for catalysis, energy storage, electronics, and much more.

## 1. Introduction

The ability to synthetically manipulate materials now surpasses what occurs in nature, down to the atomic scale. Properties central to modern technologies are ultimately rooted in a material's structure and chemistry.[1] If structure governs the available physical properties, then composition determines their magnitude. Even within a single crystal family, such as perovskites, sub-angstrom displacements can dictate whether a material is ferroelectric or paraelectric, underscoring the critical role of local structure in designing functionality. The interplay of multiple species can further enhance material performance; for example, early n-type doping of silicon with phosphorus increased its electron mobility threefold.[2,3] High entropy oxides (HEOs) extend these concepts by introducing extreme cation disorder within a crystalline framework, often maintaining high symmetry despite chemical complexity. This disorder leads to locally distinct environments that disrupt long-range periodicity, while configurational entropy expands the solubility limits of constituent elements.[4] Together, these attributes enable access to novel compositions and potentially emergent properties, making HEOs highly tunable and synthetically versatile.

The unique combination of disorder and symmetry has fueled rapid interest in HEOs, and several excellent reviews have preceded this one. Early articles focused on high entropy alloys and ceramics more broadly,[5–7] while subsequent reviews refined the scope. Musicó *et al.* provided an overview of high entropy oxides in 2020,[8] Brahlek *et al.* proposed definitions for the field in 2022,[9] and Kotsonis *et al.* linked compositional disorder with functional response.[4] Rost *et al.* offered an editorial update on recent breakthroughs that continue to expand the relevance of high entropy materials.[10]

Nature offers inspiration for such complexity as minerals exhibit remarkable chemical and structural richness, with the most complex known mineral incorporating H, C, O, Mg, Ca, and U into a tetragonal lattice.[11,12] Yet HEOs are distinct: they are synthetically designed to achieve high symmetry with extensive elemental solubility, guided by but not constrained to natural principles.[13–15] The first reported "HEO", as the community has come to accept them, $(MgCoNiCuZn)_{0.2}O$ (often referred to as "J14"), was synthesized as a rock salt structure in 2015.[16] Since then, the concept has expanded across numerous chemistries and crystallographic systems, including fluorite, pyrochlore, perovskite, spinel, garnet, and dual-phase architectures. The review on high entropy ceramics by Xiang *et al.* provides a comprehensive list of the crystal systems that were successfully synthesized as

[a.] Department of Materials Science and Engineering, Virginia Polytechnic Institute and State University, Blacksburg, Virginia 24061, USA.
[b.] Department of Materials Science and Engineering, The Pennsylvania State University, University Park, Pennsylvania 16802, USA.
† Corresponding author email: cmrost@vt.edu

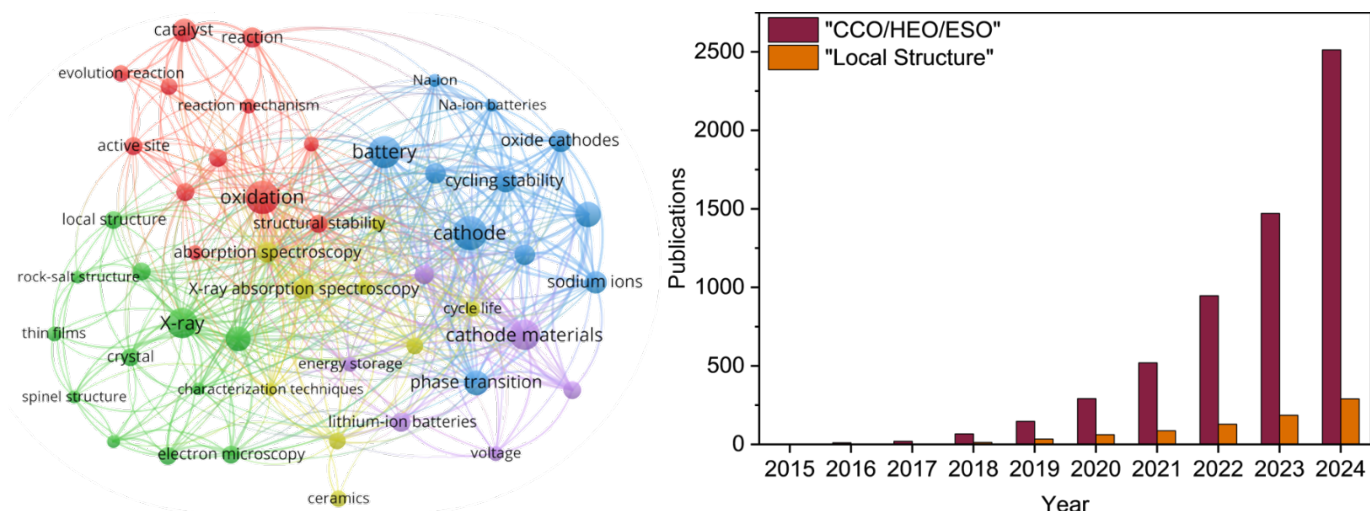

Figure 1: Concept co-occurrence network and publication trend analysis for high entropy oxide-related materials research. (Left) A concept co-occurrence map generated from literature on compositionally complex, high entropy, and entropy stabilized oxides. Node size reflects term frequency and line thickness indicated co-occurrence strength. Color-coded clusters represent thematically related research ideas, including redox chemistry (red), phase/structure characterization (green/yellow), and cathode development (blue/purple). (Right) Annual publication trends from 2015 to 2024 for two key concept searches: "compositionally complex/high-entropy/entropy-stabilized oxides" and "local structure".

high entropy compositions as of 2021.[6] Oxygen, boron, silicon, and some amounts of phosphorous, fluorine, or tellurium have been used as anions. Within the regular oxide space, rock salt, fluorite, pyrochlore, perovskite, spinel, magneto plumbite, garnet, and dual-phase systems have been created.

Though discovered nearly a decade ago, HEOs potentially rival state-of-the-art materials in several applications. From rechargeable batteries[17] to hydrogen storage[18], hydrogen evolution[19], oxygen catalysis[20], and supercapacitors[21]; high entropy compositions hold a promising future for functional materials. Figure 1 depicts a concept co-occurrence network for HEOs and synonymous materials in combination with the trends in yearly publication rates beginning with the seminal paper.[16] HEOs are of clear interest to the ceramics community. While emerging properties are still the end goal, a closer look at the local length scales in these highly disordered systems is becoming more and more necessary.

This review focuses specifically on the local structural environments in HEOs: how they arise, how they can be characterized, and their implications for material performance. We begin with foundational definitions and structural motifs, followed by a discussion of disorder and its influence on physical properties. We then summarize key characterization tools capable of probing local structure, before concluding with case studies that highlight how structural insights are informing function—and where opportunities remain for future progress.

## 2. Foundational Concepts

### 2.1 Terminology and Definitions

It is important to distinguish high entropy oxides (HEOs) from compositionally complex oxides (CCOs) and entropy-stabilized oxides (ESOs). As discussed in the review by Brahlek et al.,[9] CCOs represent the broadest category of disordered oxides and are defined as materials containing more constituent elements than their parent compositions. HEOs are a subset of CCOs, typically consisting of five or more elements in (approximately) equimolar ratios randomly dispersed across a single crystallographic site, resulting in increased configurational entropy. ESOs, in contrast, are a thermodynamic subset of HEOs where the entropy contribution is large enough to overcome a system's positive enthalpy at high temperatures, thus stabilizing a single-phase structure that would otherwise be unfavorable. Such materials are inherently metastable.

While these classifications overlap, they carry important distinctions. "High entropy" specifically refers to the configurational entropy of a system. For an ideal system with equimolar occupation of a single sublattice, the configurational entropy can be calculated as $S_{config} = R\ln(n)$, where R is the ideal gas constant and n is the number of elements on the site.[15] For five equimolar constituents, the ideal configurational entropy will be 1.609R. High entropy alloys are typically defined as having at least five components in proportions 5 % ≤ mol % ≤ 35 % or having a configurational entropy above 1.5R $(J \cdot K)/mol$.[22,23] This definition is often extended to classify HEOs.

The defining feature of ESOs is that their positive enthalpy of formation is counteracted by a sufficiently negative entropy term at high temperature, enabling stabilization of a single phase. An illustration depicting the relationships between classifications is shown in Figure 2. Rost et al. outlined four experimental criteria to demonstrate that (MgCoNiCuZn)$_{0.2}$O is an ESO.[16] They observed a reversible phase transition: a pure rock salt phase precipitated CuO upon slow cooling to 750 °C but reverted to a single-phase rock salt structure upon reheating to 1000 °C. Varying the molar ratios of cations revealed that the minimum single-phase formation temperature (875 °C) occurred at equimolar composition. Then, differential scanning calorimetry and thermogravimetric analysis confirmed an endothermic transformation, indicating that entropy drives the solid-state reaction.[24] Finally, extended X-ray absorption fine structure (EXAFS) and scanning transmission electron microscopy with energy-dispersive X-ray spectroscopy (STEM-EDS) confirmed a homogeneous cation distribution at both the local and nanometer scales.[25]

**2.2 Thermodynamic Framework**

The most common and straight forward way that high entropy systems are explained is by means of free energy relationships. Relative phase stability can be predicted by the Gibbs free energy, $\Delta G(T,P) = \Delta H(T,P) - T\Delta S(T,P)$, where $\Delta G$ is the change in Gibbs free energy $(J/mol)$, $\Delta H$ is the change in enthalpy $(J/mol)$, T is the temperature (K), $\Delta S$ is the change in entropy $((J \cdot K)/mol)$, and P is pressure (Pa). Gibbs free energy is a measure of the change in free energy of the system resulting from a process and, consequently, can predict how the system will react. The second law of thermodynamics tells us that a system will be at equilibrium when the free energy is minimized. Therefore, the Gibbs free energy equation can be used to determine the stability of a system.

In an ideal solution, the entropy of mixing is shown by a variation of Boltzmann's law, $\Delta S_{mix} = -R\sum_{i=1}^{N} x_i ln(x_i)$, where R is the ideal gas constant, $x_i$ is the molar fraction of component *i*, and N is the total number of components.[26] From this equation, it can be shown that maximum entropy of mixing can be achieved when the molar fractions are equal.

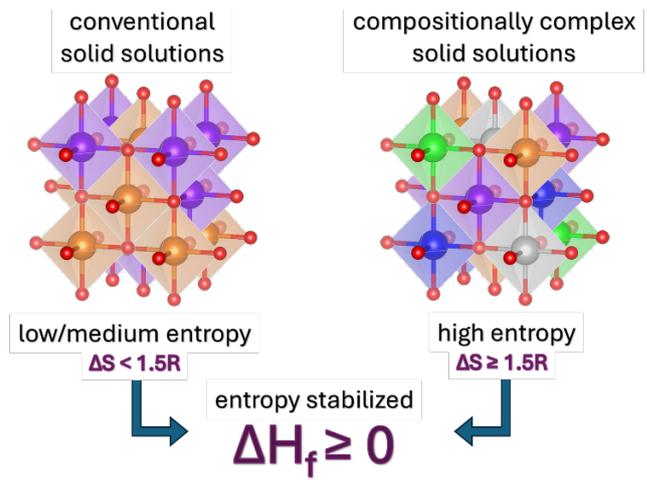

Figure 2: Simple diagram illustrating relationship between conventional and compositionally complex solid solutions. By convention, a composition is considered high entropy when the number of cation species enables ΔS ≥ 1.5R. Both compositionally complex and conventional solid solutions can be entropy stabilized if the enthalpy of formation is positive.

The enthalpy term arises in regular solutions as there is no heat associated with mixing in ideal solutions.[26] It is given in the equation $\Delta H_{mix} = \Sigma_{i,j=1}^{N} a_0 x_i x_j$, where $x_i$ and $x_j$ are the mole fraction of each component in a system with N combinations of components and $a_0$ is a constant representing the bonding preference between the constituent components. $a_0$ is unique to each combination of components $x_i$ and $x_j$ and will be positive if chemical separation is preferred and negative if alloys or intermetallic phases are preferred.[26] Substituting equations for $\Delta S_{mix}$ and $\Delta H_{mix}$ yields a more thorough equation for Gibbs free energy, $\Delta G_{mix} = \Sigma_{i,j=1}^{N} a_0 x_i x_j - RT \sum_{i=1}^{N} x_i ln(x_i)$. This equation fails to account for the fact that both enthalpy and entropy are functions of temperature and pressure, as shown before. Equations for entropy and enthalpy as functions of temperature, given that pressure is typically held constant, can be simplified to, $S(T) = S_0 + C_P \cdot ln(\frac{T}{T_0})$ and $H(T) = C_P T$, respectively, where $S_0$ is the entropy at temperature $T_0$ and $C_p$ is the heat capacity at constant pressure.[26] When these values are plotted against Gibbs free energy, it is apparent that enthalpy dominates over entropy at low temperatures but, above a sufficient temperature, entropy becomes the dominating term.

$\left(\frac{\delta G'}{\delta n_i}\right)_{T,P,n_j} = \mu_i$ tells us that the chemical potential of a component is equivalent to the partial molal free energy for the component.[26] The chemical potential of a component will determine the mixing equilibrium in a solution and is related to the activity of that component.[15] The change in chemical potential due to mixing is $\Delta\mu_i = Rln(X_i) + Rln(\gamma_i)$, wherein $\gamma_i$ is the activity of the component in solution. The $Rln(\gamma_i)$ term represents excess chemical potential stemming from the non-ideal mixing and dictates the effects of adding the component to the solution. Many partial enthalpy and partial excess entropy terms arising from different physical aspects, such as surface and magnetic effects, contribute to the excess chemical potential.[15] This tells us that non-ideal contributions to the potential gradient will have an outsized effect as there are more components intermixing in the solution.[15]

**2.3 Disorder in Oxides**

Entropy is traditionally thought of as the level of disorder in a system. While this is not incorrect, it is important to define what disorder means in the context of HEOs and think about how that produces the elevated entropy in these materials. In crystalline solids, there is no such thing as a perfect crystal, even with stoichiometrically accurate components and ideal conditions. This is due to the tendency for defects to form in order to lower the free energy of the system. In oxides, the formation of these defects is dependent on temperature, composition, and oxygen partial pressure.[27]

Intrinsic defects will form as temperature increases solely due to temperature's negative contribution to the free energy.[28] Defects in crystalline materials range in dimensionality from zero to three — point, line, planar, and volume defects. Intrinsic ionic defects manifest as point defects. Substitutional disorder is when an ion of the formal composition is substituted with another at some fraction of the sites. This can be expanded to chemical disorder if the ions across a sublattice are non-uniform. This distribution of ions will in turn lead to charge disorder as the electronic configuration varies from site to site. Increased competition for local equilibrium will also lead to the formation of point defects, which can be defined as local deviations from the long-range symmetry of the material at a single or pair of lattice sites. These can be vacancies, interstitials, Frenkel[29] and Schottky[30] pairs, or anti-site substitutions.

Differing electronic configurations inherently lead to varying bonding preferences, which in turn contribute to structural disorder. These deviations from ideal symmetry arise as ions adopt new equilibrium positions not dictated by the global periodic lattice. The resulting distortions span a range of length scales—from sub-angstrom displacements to intermediate-range order extending up to ~20 Å.[31] In studies of glasses and amorphous systems, short-range order is often divided into two categories that are also applicable to HEOs: (i) the local structural unit, and (ii) the interconnection between neighboring units.[32] In this review, we use "local structure" to refer to the former, and "short-range order" (SRO) to the latter. Jahn-Teller (J-T) distortions are a common form of local ordering in transition metal oxides.

A prominent example of local ordering in transition metal oxides is the Jahn-Teller (J–T) distortion, which involves the elongation or compression of the ligand coordination environment to lift orbital degeneracy.[33] These distortions, typically observed in ions with degenerate d-electron configurations,[34] can drive local symmetry breaking and lead to SRO depending on the population and distribution of J–T active species in the lattice.

In high entropy oxide systems, disorder emerges as a natural consequence of the thermodynamic and kinetic constraints involved in stabilizing multiple cations within a single crystallographic framework. The random distribution of chemically and electronically dissimilar elements leads to several coexisting types of disorder, including configurational, charge[35], strain and bond-length disorder[36]. Configurational disorder stems from the stochastic occupation of lattice sites by cations with varying size and valence, while charge disorder arises due to aliovalent substitutions and local charge compensation. Size mismatch among cations introduces local strain fields and lattice distortions, often accommodated through bond angle bending or polyhedral tilting. These distortions may be static or dynamic and can couple to oxygen sublattice rearrangements[37], further amplifying structural complexity. Collectively, these types of disorder are interdependent and span multiple length scales—from atomic coordination to mesoscale strain—and are central to the emergent functional properties observed in HEOs.

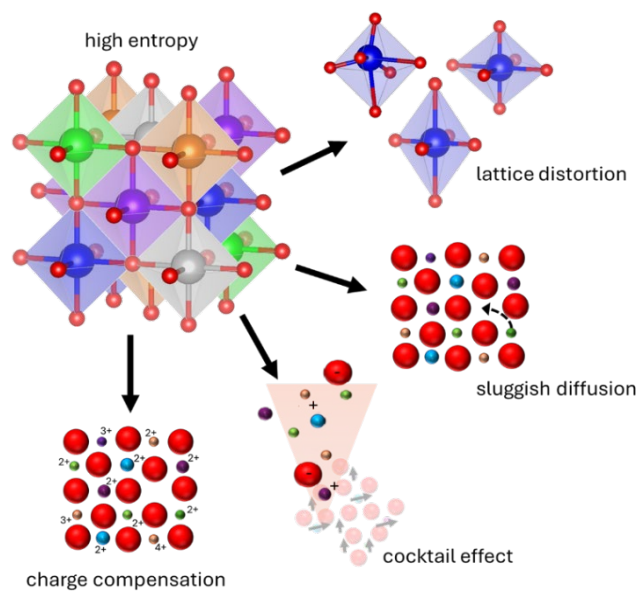

Figure 3: "4+1" core effects in high entropy oxides. The phases stabilized through high configurational entropy exhibit properties that arise from the many unique combinations of local lattice distortions and charge compensation. In turn, characteristics such as sluggish diffusion and cocktail effects may be present.

The unusual behaviors of HEOs stem directly from this intrinsic disorder—arising from the random distribution of multiple cation species across a single lattice site and the local equilibria required to maintain charge neutrality. This disruption of bonding networks and electronic environments is the fundamental driver behind many of the characteristic phenomena observed in these materials.

## 2.4 High Entropy Effects

There are four widely accepted characteristics of high entropy materials: high entropy, slow kinetics, severe lattice distortion, and a cocktail effect.[38] Research over the past two decades on high entropy materials has revealed additional defining characteristics, including enhanced stability in higher symmetry structures arising from a broad basin of attraction[39,40], structural evolution[41], progressive local structure maturation[42], and diverse functional valence environments.[43,44] Figure 3 summarizes the key topics of this section—namely the extent to which the classical characteristics can be applied to HEOs and phenomena unique to ceramic systems.

*2.4.1 High Entropy Effect* The mechanisms that lead to high entropy have been defined, but now the effect it has on the material can be discussed. HEOs contain more components on their cation sublattice than traditional oxides. The high configurational entropy induced by the chemical disorder on one (or more) sites of the compound allows for new compositional space to be explored. Typically, this is achieved through incorporating many constituent elements into the material. However, it has been shown that entropy stabilization can occur when the configurational disorder on the cation lattice alone does not reach the 1.5R $(J \cdot K)/mol$ threshold. An oxide was synthesized that only had 1.24R $(J \cdot K)/mol$ of entropy when considering the cation sublattice, but was stabilized due to vacancies on the oxygen sublattice increasing the configurational entropy.[45] There has been work done in chemically disordering the anion sublattice as well[46–54], but this departs from the realm of oxides and is therefore omitted from this review.

In high entropy alloys, there have been arguments against configurational entropy being the true force behind the single-phase stabilization of these materials.[7,55,56] Citing the lack of single-phase chemical homogeneity observed in materials many element alloys, the varying enthalpies for intermetallic phases, and loss of single-phase nature when replacing select elements; there is a valid argument to be made. However, there are computational and experimental studies that display entropy being the driving factor in phase stability in high entropy materials, highlighting the necessity of high temperatures in the synthesis to scale the entropic contribution.[16,57–59]

*2.4.2 Severe lattice Distortion* A consequence of substituting differing atomic radii across a lattice site will result in anisotropic bond lengths within the material. This impacts more that the structure alone, as Guan *et al.* demonstrated that the shortening of metal-oxygen bonds in a high entropy spinel led to improved stability and performance in catalytic applications.[36] (MgCoNiCuZn)$_{0.2}$O contains an near-ideal cation matrix, with structural compensation occurring through distortions existing on the oxygen sublattice using EXAFS measurements which agreed with density functional theory (DFT) calculations.[60] Octahedrally coordinated Cu is J-T active and, as such, results in additional lattice distortions around Cu populated sites.[61–63] Rák *et al.* found that Cu being in the rock salt structure resulted in an abnormally high proportion of J-T compression, as opposed to J-T elongation, possibly due to structural competition between elongation preferred by Cu and octahedral structures preferred by the other cations in the lattice.[64]

***2.4.3 Sluggish Diffusion*** The driving force for diffusion is directly related to the chemical potential gradient.[65] The homogenous distribution of cations in HEOs results in an average chemical potential gradient that is flatter than would occur in conventional oxides, reducing the drive for mass transport. While the average potential gradient is flatter, there is still the opportunity for local deviations to be large and form 'traps' for migrating ions. Lattice distortions, misaligned lattice sites, and different ionic radii create physical barriers that can stunt the movement of ions, contributing to slow diffusion. Slow diffusion characteristics of HEOs are observed in mechanisms like vacancy migration, interstitial diffusion, and grain growth. In a high entropy environment, the creation and movement of cation vacancies is often more complex due to the diversity of electronegativity and atomic radius of the constituent elements. Interstitial diffusion is stunted in HEOs as the diverse potential landscape limits the number of suitable interstitial sites and increases the activation energy for diffusion[66]. Sluggish diffusion effects have been observed in the reduction of grain growth rates as well[67], which can be beneficial in high temperature applications.[68] It is important to note that the concept of sluggish diffusion being an intrinsic characteristic of high entropy materials has been contested, especially in alloys where fundamental diffusion studies are more prevalent.[69,70]

***2.4.4 Cocktail Effects*** The cocktail effect is something that is not well understood but is observed via many of the surprising properties that arise. It is best described as the synergistic combination of the properties of the constituent elements, their interactions, and the unique structure of the system that results in behavior that cannot be ascribed to any one component. The excess chemical potential, discussed in section 2.2, for each component in the solution is thought to be the thermodynamic origin of these phenomena.[15] The term came about in Ranganathan's discussion of complex alloys[71] and was realized to be a core property of high entropy alloys.[38] The concept is apparent in HEOs as well, with evidence in catalytic applications being especially strong.[72] For example, Hao *et al.* created high entropy perovskite nanofibers for the purpose of alkaline oxygen evolution reaction. They found that their inclusion of Cu on the B-site out-performed perovskites of similar composition, as well as the commercially available standards.[73]

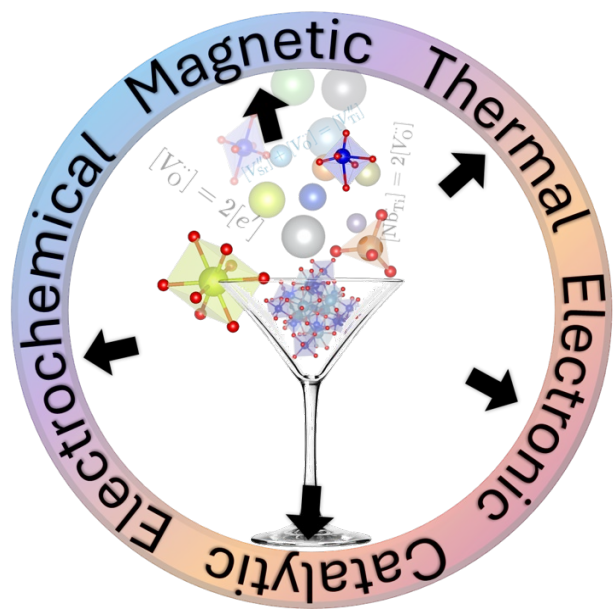

Figure 4: The disorder cocktail in HEOs. Illustration depicting the types of local disorder including but not limited to point defects, cation charge and size variation, and polyhedral distortions, that may co-exist in any HEO system and the material properties they have been shown to impact.

***2.4.5 Charge Compensation*** A particularly intriguing trait of HEOs is the charge compensation effect. In ceramics, the cations on a particular site ideally have a uniform valence directly balancing the anions to achieve charge neutrality. However, when elements with differing electronic configurations are present, there are two possible mechanisms through which balance can be achieved. Firstly, a neighboring atom can undergo oxidation to account for the charge imbalance. For example, if $Li^+$ replaces Ni in a $Ni^{2+}$ matrix, a neighboring Ni atom will oxidize and become $Ni^{3+}$ in order to balance the charge of the local system.[74] Alternatively, charge compensation occurs through the formation of vacancies on the oxygen sublattice. Using the previous example of lithium in a nickel matrix, the introduction of an oxygen vacancy, $V_O^{\bullet\bullet}$, can offset two $Li^+$ substitutions.[74] The capability to substitute aliovalent atoms into a HEO lattice was first demonstrated by Bérardan *et al.* wherein elements of $1^+$, $3^+$, and $4^+$ valence states were substituted into $(MgCoNiCuZn)_{0.2}O$.[57] This study showed that a single-phase rock salt structure was still formed with the addition of lithium up to equimolar proportions in a six-component material. Measuring the lattice parameter of this composition demonstrated compliance with Vegard's law, indicating Li was indeed substituted into the structure rather than occupying interstitial sites. The decrease in lattice parameter additionally suggests that charge compensation mechanisms occurred, either through oxidation of other cations or through formation of oxygen vacancies, as Li alone should increase the lattice parameter. Further, it was shown that the addition of $3^+$ valence elements resulted in the formation of secondary phases whereas addition of equal proportions of $3^+$ and $1^+$ ions resulted in a single phase.

Table 1: Summary of average versus local structural features in key high entropy oxide (HEO) crystal systems and their implications for functional properties. While the average crystallographic structure determines long-range order, local chemical and structural disorder—driven by cation size/valence mismatch and oxygen sublattice distortions—plays a critical role in governing transport, ferroic, magnetic, and catalytic behavior. References are provided to guide further exploration of structure–property relationships across diverse HEO systems.

| 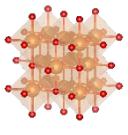 | 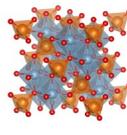 | 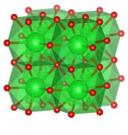 | 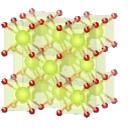 | 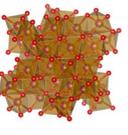 | 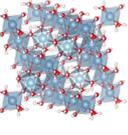 | 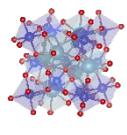 | 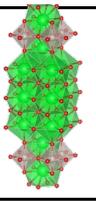 |
|---|---|---|---|---|---|---|---|
| rock salt (Fm-3m) | spinel (Fd-3m) | perovskite (Pm-3m) | fluorite (Fm-3m) | bixbyite (Ia-3) | pyrochlore (Fd-3m) | garnet (Ia-3d) | Ruddlesden-Popper (I4/mmm) |
| Cations occupy octahedral sites in a cubic lattice. | A-site (tetrahedral), B-site (octahedral); normal or inverse. | Ideal cubic or distorted $ABO_3$ lattice. | FCC with fully occupied cation lattice, disordered O sublattice. | Defective fluorite; two inequivalent cation sites, intrinsic O vacancies. | Ordered $A_2B_2O_7$ with corner-sharing $BO_6$ octahedra. | Complex framework of dodecahedral, octahedral, and tetrahedral sites. | Alternating perovskite and rock salt layers. |
| Strong local distortions due to size/valence mismatch; site-specific bond length variations. | Cation redistribution and short-range order (SRO); B-site cation disorder common. | Octahedral tilting, local symmetry breaking; polar/non-polar domain coexistence in relaxors. | Strong local oxygen vacancy clustering, asymmetric cation-O environments. | Local distortions due to cation size/charge mismatch, vacancy clustering, potential site preference. | Local disorder on A and B sites; O' disorder affects $BO_6$ connectivity. | Uneven local coordination geometries; site-specific bonding variations. | Interlayer strain, stacking faults, B-site order/disorder. |
| Affects thermal conductivity, electrical transport, and phonon scattering. | Alters redox activity, magnetism, and battery performance. | Enhances pyroelectric/ferroelectric response, local dipoles affect dielectric behavior. | Impacts oxygen ion mobility and redox cycling; critical for fuel cells and catalysis. | Influences optical transparency, ionic conductivity, and electronic structure. | Affects magnetic frustration, ionic conduction, and thermal robustness. | Critical for $Li^+$ transport in solid electrolytes; local distortions modulate conductivity. | Enables anisotropic ion/electron transport, relevant for thermochromics and cathodes. |
| 57,63,75–107 | 108–115 | 116–126 | 127–133 | 134–136 | 137–142 | 143 | 144–146 |

## 3. Local Structure-Property Relationships

The compositional complexity and associated "disorder cocktail" in HEOs give rise to physical properties that diverge from those of conventional oxides. The influence of local structure on thermal, electronic, and magnetic behavior is particularly pronounced in HEOs due to their compositional complexity and structural flexibility. Table 1 provides a comparative summary of how various types of disorder impact these properties across representative systems. One hallmark is enhanced thermal and structural stability at elevated temperatures, a consequence of entropy-driven phase stabilization. Conversely, ESOs are prone to phase separation at lower annealing temperatures, where the entropic contribution to the free energy becomes insufficient to counter enthalpic driving forces.[147] These fundamental thermodynamic distinctions lay the groundwork for a wide spectrum of unique and tunable properties in HEOs. Thermal transport can be dramatically suppressed due to complex phonon scattering mechanisms, while ionic conductivity and catalytic activity are often enhanced by defect-rich, disordered environments. Magnetic behavior ranges from long-range order to spin-glass states depending on compositional tuning, and electrochemical stability can be achieved through selective redox participation of constituent ions. These examples underscore how configurational disorder not only modifies conventional property trends but also creates new opportunities for multifunctionality. The following highlights some of the examples of properties observed in various HEO compositions.

## 3.1 Thermal Conductivity

Thermal transport in HEOs is governed by complex phonon scattering mechanisms that go beyond conventional mass disorder. Recent studies have shown that charge disorder, lattice distortions, and compositional complexity can dramatically suppress thermal conductivity, offering new design strategies for thermoelectric and insulating materials.

Braun *et al.* investigated the thermal conductivity of $(MgCoNiCuZn)_{0.2}O$ and its six-component derivatives, finding that the observed ultralow values could not be explained by phonon scattering from mass disorder alone[80] Instead, their analysis pointed to Rayleigh-type scattering caused by large fluctuations in interatomic force constants originating from charge disorder, and subsequent localized polyhedral disorder, as a dominant mechanism. This deviation from conventional phonon scattering models also led to an atypical decoupling between thermal conductivity and stiffness: despite strong suppression of heat transport, the elastic moduli remained high, violating the typical correlation observed in non-metals.

Jana *et al.* extended this understanding to a novel high entropy niobate, $(SrBaLiKNa)_{0.2}Nb_2O_6$, which crystallized in a tungsten bronze structure and was designed for thermoelectric applications.[148] The measured lattice thermal conductivity approached the theoretical minimum (~0.7 W m$^{-1}$ K$^{-1}$), with a value of 0.8 W m$^{-1}$ K$^{-1}$ at 470 K—indicative of glass-like thermal transport dominated by phonon scattering from point defects and structural disorder. Minimal electronic contribution confirmed that heat conduction was governed primarily by lattice vibrations.

More recently, Riffe *et al.* demonstrated that broadband optical phonon scattering, intensified by compositional complexity and local lattice distortions, plays a central role in reducing thermal conductivity in multi-cation oxides.[149]. Their combined experimental and computational study showed that diverse cation environments disrupt phonon lifetimes—particularly through optical modes—providing a robust framework for tuning thermal transport beyond mass contrast effects alone.

## 3.2 Electrochemical and Catalytic Activity

HEOs exhibit promising performance in both electrochemical and catalytic applications, owing to their structural stability, defect tolerance, and tunable redox behavior. In the electrochemical realm, doped variants of $(MgCoNiCuZn)_{0.2}O$ have shown high ionic conductivity for lithium and sodium, highlighting their potential for use in all-solid-state batteries.[150,151] This conductivity was attributed to the migration of ions through oxygen vacancies formed via charge compensation upon doping.[151] Additionally, $(MgCoNiCuZn)_{0.2}O$ has demonstrated excellent capacity retention as a lithium-ion battery anode.[96,152] This is primarily due to partial redox participation where only a subset of cations undergo lithiation/delithiation. The remaining elements help preserve the rock salt structure and mitigate structural degradation during cycling.[96,152]

In catalytic contexts, similar structural features, namely lattice distortions and configurational disorder, also play a role. Zhang *et al.* found that oxygen vacancy formation energies were significantly reduced in high entropy compositions, potentially due to weakened Coulombic attraction between cations and anions.[153] Their HEO exhibited a greater concentration of oxygen vacancies than binary or ternary counterparts, resulting in enhanced combustion catalysis of $C_3H_6$. Likewise, Li *et al.* reported that $CuZnAl_{0.5}Ce_5Zr_{0.5}O$ outperformed simpler oxides in the $CO_2$ hydrogenation reaction and demonstrated improved catalytic durability over 100 hours.[154] These examples underscore the ability of HEOs to combine chemical complexity with functional robustness, enabling performance advantages across both ionic transport and redox-driven catalytic processes.

## 3.3 Magnetism

There is a diverse range of magnetic behaviors that are highly sensitive to composition, structure, and site-specific interactions in HEO systems. $MgCoNiCuZn)_{0.2}O$ demonstrates long-range antiferromagnetic ordering, despite containing only a subset of magnetic ions—Ni, Co, and Cu.[62,155] Reducing the concentration of these magnetic ions through chemical substitution alters the ground state, leading to spin-glass behavior, and subtle changes in composition have also been shown to influence the effective magnetic moment and the magnetic transition temperature.[155] In particular, J-T distortions modulated by the Cu content have been used to tune the degree of magnetic frustration and glassiness in the system.[61,156]

Table 2: Summary of characterization techniques used to probe local and atomic-scale structure in materials science. Techniques are organized by the characteristic length scale probed, highlighting key features and types of structural or electronic information obtained. The rightmost column provides representative reference ranges for publications utilizing each method.

| Technique | Length Scale | Key Features | Information Gained | Publications |
|---|---|---|---|---|
| X-ray Absorption Fine Structure (XAFS) | 2–6 Å | Element-specific, probes oxidation state and coordination | Coordination number, bond lengths, oxidation states | 160,163–179 |
| Pair Distribution Function (PDF) | ~2–30 Å | Captures short- and medium-range order including amorphous phases | Atomic pair correlations in real space | 180–186 |
| Transmission Electron Microscopy (TEM) | 0.1–10 Å | Atomic-resolution imaging; interface/defect analysis | Atomic positions, defects, grain boundaries | 187–199 |
| Electron Energy Loss Spectroscopy (EELS) | ~1–2 Å | Elemental & bonding information with atomic spatial resolution | Valence states, coordination, bonding | 200–209 |
| Atom Probe Tomography (APT) | Sub-nm (x, y); few nm (z) | 3D atom-by-atom elemental mapping | Elemental identity and spatial distribution | 210–214 |
| Mössbauer Spectroscopy | Atomic-scale | Sensitive to hyperfine interactions; element-specific | Magnetic/electronic environment at nuclei | 215–223 |
| Nuclear Magnetic Resonance (NMR) | Atomic-scale | Identifies local bonding & structure; useful in disordered systems | Chemical environments, short-range structure | 224,225 |
| Raman Spectroscopy | Micron-scale (optical); vibrationally local | Non-destructive, bond- and phase-sensitive; suitable for mapping | Vibrational modes, bonding, symmetry, disorder, strain | 226–233 |

In perovskite-structured HEOs, magnetism is largely governed by B-site cations due to their direct role in mediating super exchange interactions.[61,157,158] However, A-site contributions become increasingly relevant at lower temperatures, where rare-earth elements can introduce localized magnetic interactions. Importantly, studies have revealed a clear relationship between the ionic radius of A-site cations and the Néel temperature, mediated through the Goldschmidt tolerance factor and its influence on the B–O–B bond angle.[157,158]

Krysko *et al.* examined the magnetic behavior of a single crystal high entropy aluminate spinel, $(MgMnFeCoNi)_{0.2}Al_2O_4$, and identified a cluster spin glass phase, evidenced by bifurcation in field-cooled, zero-field-cooled magnetization measurements[159]. This magnetic ground state reflects a combination of spin glass and antiferromagnetic behavior seen in the parent compounds, influenced by severe lattice distortion and magnetic frustration. Single crystalline high entropy spinel ferrites $(Mg_{0.2}Mn_{0.2}Fe_{0.2}Co_{0.2}Ni_{0.2})_xFe_{3-x}O_4$ were shown to exhibit sharp ferrimagnetic transitions up to 748 K, despite substantial cation disorder[160]. EXAFS analysis suggested a random cation distribution across lattice sites, which suppressed local clustering and short-range order up to x = 1.8, thereby enhancing magnetic homogeneity and preserving transition sharpness. These examples highlight how cation disorder and competing magnetic interactions in high entropy spinels give rise to complex magnetic ordering.

Overall, while the magnetic properties of HEOs often reflect those of their magnetic constituents, the ability to systematically vary composition enables precise tuning of magnetic ground states and transition temperatures. Such tuning offers a versatile platform for exploring frustration, glassiness, and emergent magnetic order.

### 3.4 Electronic Properties

Several studies have explored the electrical properties of HEOs, revealing behaviors distinct from those of conventional ceramics. One early report identified a colossal dielectric constant (> $10^5$ at 373 K and 20 Hz) in $(MgCoNiCuZn)_{0.2}O$, suggesting significant potential for dielectric applications.[57] High entropy strategies applied to potassium sodium niobate (KNN) ceramics have yielded enhanced breakdown fields and high energy storage efficiency, attributed to grain refinement and the presence of polar nano regions.[161,162] Liu *et al.* observed high piezoelectric coefficients in high entropy perovskites, which correlated with increasing configurational entropy and were attributed to a flexible polarization configuration driven by lattice distortions.[234] Uniquely, the high entropy perovskites were shown to have a flexible polarization configuration stemming from lattice distortions. In addition to dielectric and ferroelectric properties, band gap tunability has been demonstrated in entropy-stabilized systems: Sarkar *et al.* used Pr substitution to modify the band gap in medium-entropy rare earth oxides,[235] while Chen *et al.* achieved both band gap tuning and improved breakdown strength in KNN-based HEOs through Hf and Ta doping.[162]

Uncovering the remarkable capabilities of HEOs holds great promise for the future of materials research. However, to fully exploit their tunability, it is essential to understand the underlying mechanisms by resolving the disorder and local structure features within these systems.

## 4. Characterization of Local Disorder in HEOs

Characterizing HEOs is both fundamentally necessary and greatly challenging. While macroscopic techniques such as XRD and SEM-EDS confirm the bulk symmetry and average elemental distribution of the system, they do not complete the picture. In HEOs, lattice distortions and defects often occur within the single-phase regime, and it is essential that materials be randomly homogenous at the atomic scale to maximize configurational entropy. Therefore, special emphasis must be placed on investigating the local structure, as it is this fine-scale arrangement that gives rise to many of the unique properties observed in HEOs. Given the distribution of cations across a single lattice site and the resulting angstrom-scale distortions, advanced techniques with high structural resolution and elemental selectivity are critical. This section highlights some of the premier characterization methods that fulfill these requirements, a more comprehensive list of techniques and associated references are in Table 2.

### 4.1 Spectroscopy

Spectroscopic techniques offer critical insight into the local structural, electronic, and magnetic environments of high-entropy oxides (HEOs), complementing average structural data obtained by diffraction. These methods are particularly powerful in assessing cation valence states, site symmetries, bonding environments, and magnetic interactions—features often obscured in disordered, compositionally complex systems. Figure 4 provides a comparative overview of the primary spectroscopic methods discussed in this section. We highlight the utility of X-ray absorption fine structure (XAFS), X-ray magnetic circular dichroism (XMCD),

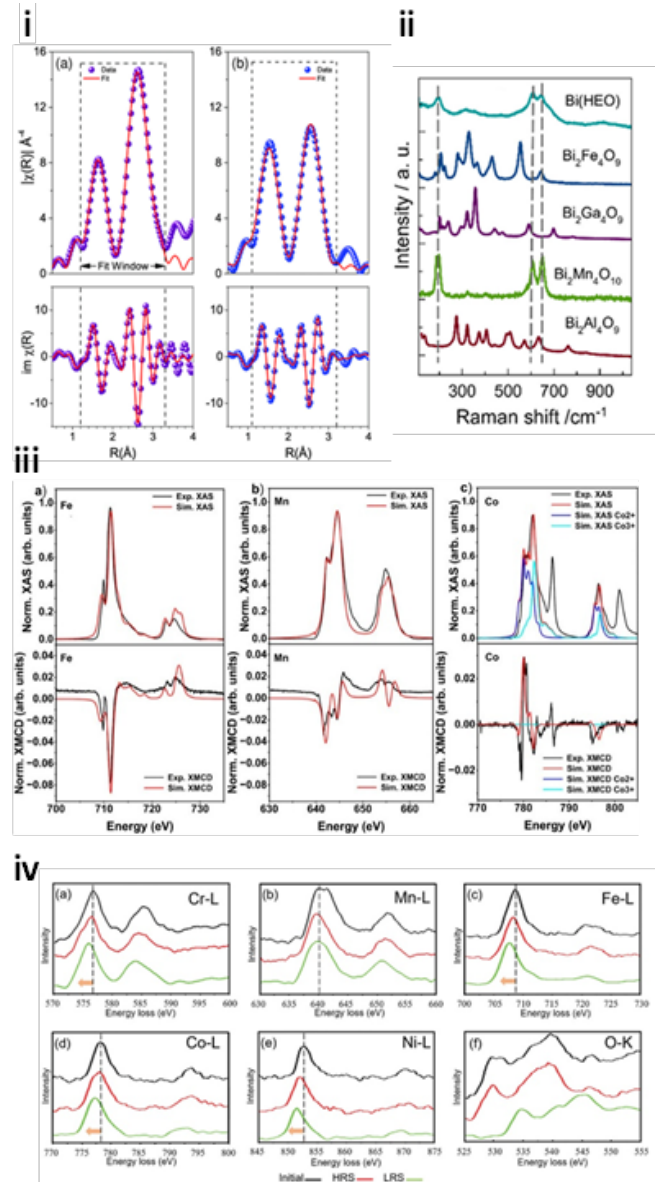

Figure 5 (i) Extended X-ray absorption fine structure (EXAFS) data and best-fit models for the Co K-edge in (MgCoNiCuZn)$_{0.2}$O and (MgCrCoNiCuZn)$_{0.167}$O. Both the magnitude and imaginary components of the real-space function χ(R) are shown as a function of distance R from the cobalt absorber. These fits reveal differences in local coordination environments influenced by the addition of Cr, consistent with increased structural disorder. Reproduced from Braun, *et al.* Adv. Mater. 2018, 30, 1805004. (ii) Raman spectra of Bi$_2$(Al$_{0.25}$Ga$_{0.25}$Fe$_{0.25}$Mn$_{0.25}$)$_4$O$_9$ compared with parent compounds Bi$_2$Al$_4$O$_9$, Bi$_2$Ga$_4$O$_9$, Bi$_2$Fe$_4$O$_9$, and Bi$_2$Mn$_4$O$_{10}$. Spectral shifts and broadening indicate the effect of cation mixing on vibrational modes. Adapted under CC-BY-NC-ND 4.0 from Kirsch, *et al.* Chemistry of Materials 2023 *35* (20), 8664-8674. (iii) X-ray absorption (XAS) and corresponding x-ray magnetic circular dichroism (XMCD) spectra for Fe, Mn, and Co (top and bottom panels, respectively), with experimental data and CTM4XAS fits recorded at 2.5 K. Adapted under CC BY from Regmi *et al.* AIP Advances 14, 025023 (2024). (iv) Electron energy loss spectroscopy (EELS) analysis of a Pt/(CrMnFeCoNi)$_3$O$_4$/Nb:STO resistive RAM device at different resistance states. Spectra for the Cr L-edge (a), Mn L-edge (b), Fe L-edge (c), Co K-edge (d), Ni K-edge (e), and O K-edge (f) show state-dependent changes in electronic structure. Black, red, and green lines correspond to the initial, high-resistance, and low-resistance states, respectively. Reproduced with permission from J.-Y. Tsai *et al.*, *Adv. Mater.* 2023, 35, 2302979. Copyright 2023, Wiley.

Raman spectroscopy, electron energy loss spectroscopy (EELS), and Mössbauer spectroscopy in probing local structure and chemical behavior in HEOs.

*4.1.1 X-Ray Absorption Fine Structure.* X-ray absorption fine structure (XAFS) is an extremely powerful technique for exploring the local structure of HEOs as it is sensitive to both the species and local environment of selected absorbers. XAFS is a spectroscopic technique that returns the absorption coefficient as a function of energy $\mu(E)$ through the Beer-Lambert law, $I_t = I_0 e^{-\mu x}$ or $ln\left(\frac{I_0}{I_t}\right) = \mu(E)x$, where $I_t$ and $I_0$ are the intensity of the transmitted and incident beams, respectively, and x is the sample thickness.

XAFS spectra is divided into two regions based on the source of the signal, XANES and EXAFS, which is delineated by the absorption edge. This spectral feature is the sharp increase in the absorption coefficient associated with the energetic activation of the X-ray absorption mechanisms. The absorption edge is what allows this technique to be element specific as it loosely corresponds to the mass of the nucleus. The edge energy is sensitive to the charge state of the absorber, where the energy will increase as the ion's charge state increases. It has been shown that when compositions have the same species of ligand, this edge energy has a linear relation to the charge state.[236]

X-ray absorption near edge structure (XANES) is the region from around 100 eV before the edge to 30-50 eV above the absorption edge. The information in this region corresponds to the unoccupied states of the absorber. In addition to charge state, the XANES region can inform on the coordination environment through pre-edge peaks or shoulders in the spectra.[237]

Extended X-ray absorption fine structure (EXAFS) is from 30-50 eV after the absorption edge to continuum and this is the signal that results from the photoelectron scattering off the local environment and back to the absorber. The superposition of these interference waves is reported in reciprocal space by wavenumber (Å$^{-1}$) and is represented by the EXAFS equation, $\chi(k) = \sum_i \frac{N_i S_0^2}{kR_i^2} f_i(\pi, k) e^{(-2\sigma_i^2 k^2)} e^{(-\frac{2R_i}{\lambda_i(k)})} \sin(2kR_i + \delta_c(k) + \delta_i(k))$. This data is not easily interpreted and is Fourier transformed into a partial-pair distribution function reported in real space (Å). The EXAFS region informs on the bond lengths, coordination, and geometry of the nearest and next-nearest neighbors of the absorber.

The analysis of XAFS data differs between the two regions. For EXAFS, data is modelled by an equation, which allows structural models to be fit by refining select parameters from the EXAFS equation—namely the path lengths, degeneracy of paths, and the Debye-Waller factor. The nuances of XANES data limit analysis to qualitative techniques. Fingerprinting is the most straightforward technique used where either calculated or measured standards are compared to experimental data to match spectral features corresponding to known differences.[238] The Demeter package of programs is the premier software for processing and analyzing XAFS data, with Athena used for processing and Artemis for EXAFS analysis.[239] These programs are coded in PERL and are unfortunately no longer supported, but Larch is written in Python and is actively being improved.[240]

XAFS has been leveraged to investigate the local structure of HEOs since their inception. *al.* used EXAFS to support the claim that the cations occupied the same site within rock salt in (MgCoNiCuZn)$_{0.2}$O by comparing the partial pair distances for each absorber.[25] Subsequent studies have used XAFS to indicate charge compensation[241], Jahn-Teller distortions[60], and coordination[242]. As will be seen in the case studies, XAFS is used frequently in conjunction with other, complementary characterization techniques.

*4.1.2 X-ray Magnetic Circular Dichroism* XMCD is a variant of X-ray absorption spectroscopy (XAS) that uses left- and right-circularly polarized light to probe the electronic and magnetic structure of materials.[243] The XMCD spectra is the difference of the left and right polarized scans. This technique retains the benefits of chemical and site sensitivity that is offered by XAFS through resonance energy and spectral features. Soft X-ray edges are advantageous to hard edges for XMCD as they drive dipole-allowed transitions to the magnetically active valence band. In a *3d* transition metal, the core *2p* state is split by spin-orbit interactions into spin 3/2 and 1/2 levels. The helicity of the incident light selectively excites electrons from one of these bands, corresponding to the *L$_2$* and *L$_3$* edges. The available holes in the *3d* band dictate the sign of the signal that is output.[243] This translates accordingly to appropriate edges for elements with more complex electronic configurations. The dichroism can further inform on the site selectivity in materials. For example, the tetrahedral and octahedral sites in ferrite spinels have opposite magnetic alignments.[243] There are sum rules that allow the integral of the XMCD signal to be correlated to the orbital and spin magnetic moments.[244,245] This technique has been applied to HEO thin films to extract the magnetic orientation and interaction between elements within perovskite and spinel crystal systems.[246–248]

*4.1.2 Raman Spectroscopy*. Raman spectroscopy allows examination of local structure by probing the vibrational modes of a material.[249] When monochromatic IR light interacts with a sample, it scatters and shifts in frequency depending on the vibrational modes of the material which are influenced by atomic arrangement, local bonding, and symmetry.[250] The frequency shift can be positive or negative, depending on whether the incident light gains or loses energy upon interaction with the sample. By analyzing these frequency shifts in terms of wavenumber ($cm^{-1}$), Raman spectra are obtained, with specific peaks corresponding to the different vibrational modes.[250] A positive shift in wavenumber, known as "Stokes scattering," occurs when the material is excited to a higher vibrational state, reducing the energy of the scattered light. Conversely, a negative shift in wavenumber, called "Anti-Stokes scattering," occurs when the material relaxes to a lower vibrational mode, increasing the energy of the scattered light.[249] If the scattered light retains the same wavelength as the incident light, it is known as Rayleigh scattering.[249]

In HEOs, the vibration of metal-oxygen bonds contributes to the distinct peaks in the Raman spectra.[251] Variation in species and bond lengths throughout the material causes dispersion of energies for vibrational modes, resulting in broadening and shifting of Raman peaks. The width and position of these altered peaks can be used to characterize the degree of disorder associated with specific bonds in the structure.[252] Raman spectroscopy has been used to help characterize the homogeneity in HEOs.[251] Raman spectroscopy has also been used to help determine the presence of oxygen vacancies in (CeLaPrSmY)O HEOs as a function of pressure.[253]

*4.1.3 Electron Energy Loss Spectroscopy.* Electron energy loss spectroscopy (EELS) is commonly used in conjunction with electron microscopy to provide information about elemental composition, chemical bonding, and electronic structure of materials.[254] It is essentially the electron-based analog to XANES. This technique works by analyzing the energy loss of electrons as they pass through a sample, which gives insight into materials electronic structure and local environment.[254] EELS is typically performed in TEM or STEM setups, enabling it to be coupled with direct imaging techniques to gain complementary information about the sample. When the electron beam interacts with the sample, some electrons lose energy through inelastic scattering, leading to specific excitations within the material. These energy losses can be measured and used to extract valuable information regarding various material features, such as elemental composition, oxidation states, and bonding characteristics.[254,255]

EELS can be divided into two regions: low-loss and core-loss. These regions provide different types of information.[254] The low-loss region (typically from 0 to ~100 eV) primarily provides information on the electronic structure of the material, such as the bandgap, plasmons, and inter-band transitions. This region can be useful for understanding the material's electronic properties but does not give direct elemental or chemical composition details. On the other hand, the core-loss region (typically above 100 eV) is focused on elemental and chemical analysis. In this region, electrons are excited from core-level orbitals to unoccupied states, which corresponds to characteristic absorption edges for different elements.[254] The *K*-edge and *L*-edge are often used to identify specific elements and their oxidation states. However, one limitation of EELS is its reduced sensitivity to lighter elements like oxygen, which makes analyzing these elements challenging.

EELS has been used by Ayyagari *et al.* to observe local variations in Mn and Co valency in a HEO thin film along its thickness.[256] Phakatkar *et al.* employed EELS to determine the elemental homogeneity in a six-component HEO nanoparticle.[257] Additionally, Kante *et al.* used EELS to analyze the oxidation states of constituent elements in HEOs, such as Cr, Mn, Fe, Co, and La, by examining the edge positions and intensity ratios of $L_2/L_3$ and $M_4/M_5$ edges, providing insights into charge transfer and the preferred valence states of transition metals and lanthanides.[258] EELS analysis of an HEO particle by Baek *et al.* showed varying oxidation states across its surface and bulk, with $Fe^{2+}$ and $Co^{3+}$ more prevalent on the surface, and $Fe^{3+}$ and $Co^{2+}$ toward the center.[259] The O *K*-edge data indicated higher oxidation states at the surface, likely due to mixed oxidation states across the particle.[259]

*4.1.4 Mössbauer Spectroscopy* The Mössbauer effect is the recoilless emission and resonant absorption of gamma rays (γ) by nuclei fixed in a solid.[260] When the γ source is oscillated to create a spectrum of photon energies, the resulting absorption data can provide information on the local electronic and magnetic environments of select atomic species. The γ source is an unstable isotope of Z+1 to the targeted (Mössbauer active) isotope so that the when the source isotope decays, the emitted photons are resonant with the isotope in the sample and will thus be absorbed. In order for isotopes to be Mössbauer active, the γ energy needs to sufficient low for recoilless absorption and the lifetime of the excited state upon absorption needs to be relatively long to increase the resolution of the measurements.[261] There are several isotopes that are Mössbauer active, but $^{57}$Fe is the most commonly probed due to its natural occurrence.[262] $^{57}$Fe has γ energy of 14.4 keV and an excited state lifetime of 141 ns, correlating to a line width of $4.67 \times 10^{-9}$ eV, which facilitates highly precise measurements.[261] The mechanism by which Mössbauer spectroscopy provides information is hyperfine interactions where the electron cloud surrounding the nucleus splits the nuclear levels. There are three major results of these hyperfine interactions that appear in the spectra and inform on the local environment of the target isotope.

Electric monopole interactions between protons and *s*-electrons that penetrate the nuclear field cause a shift in the observed spectra known as the isomer shift (δ). This shift is sensitive to the density and radial distribution of *s*-electrons, and it provides information on the valence state and chemical bonding of the isotope. Electric quadrupole interactions appear in the spectra as split resonance lines, with a spectral separation directly corresponding to the energy between nuclear substates. These interactions occur when nuclear states possessing a quadrupole moment are exposed to a non-zero electric field gradient. The electric field gradient arises from lattice asymmetry and anisotropic valence electron distributions, allowing the quadrupole splitting to inform on oxidation state, spin state, and local symmetry. In $^{57}$Fe, this manifests in the spectra as a doublet. Magnetic dipole interactions occur when the nucleus possesses a magnetic moment and experiences a local magnetic field. The field lifts the degeneracy of the nuclear spin states and results in splitting into $2I+1$ substates, where $I$ is spin. The observed hyperfine splitting informs on the magnitude of the local magnetic field, which is caused by *s*-electron spin density at the nucleus, orbital motion of valence electrons, and/or total electron spin. This provides information about magnetic ordering and electron spin density in the material. In $^{57}$Fe, these interactions manifest as a sextet of resonance lines, resulting from the excited state splitting into four sublevels and the ground state into two.[261]

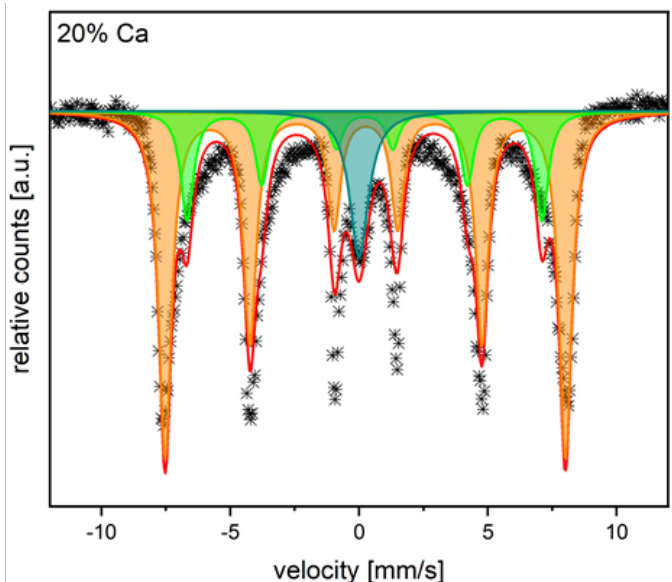

Figure 6: Fitted Mössbauer spectrum of 20% Ca in $(Gd_{0.2}La_{0.2}Nd_{0.2}Sm_{0.2}Y_{0.2})_{1-x}Ca_xFeO_3$ using two magnetic sextet models corresponding to $Fe^{3+}$ and one singlet for $Fe^{4+}$. Reproduced through CC-BY-NC-ND from Eiselt *et al.* Int J Appl Ceram Technol. 2023; 20: 213–223.

This technique is extremely powerful for probing the electronic configuration and local magnetic environment of select isotopes with high precision. For HEOs containing Mössbauer active nuclei, it yields valuable insights on the local structure in such complex environments. It has been used to elucidate the site occupancy of Fe in spinels[215,219–221], confirm magnetic ordering temperatures[215,216,220], and determine oxidation state of Fe.[216,217,219] Eiselt *et al.* was able to discern magneto-electric phase separation in crystallographic single-phase high entropy perovskite through the presence of two different $Fe^{3+}$ environments as well as $Fe^{4+}$ compensating for vacancy formation, shown in Figure 6.[216] Sarkar *et al.* discovered Néel-type collinear spin arrangement and site occupancy of Fe in an inverse high entropy spinel.[220]

**4.2 Electron Microscopy**

Electron microscopy (EM), particularly scanning transmission electron microscopy (STEM), is a powerful imaging technique capable of resolving complex atomic structures and compositional variations in HEOs at sub-nanometer scales. Traditional methods like XRD provide average structural information but cannot reveal the subtle local ordering, nanoscale variations, or chemical segregations critical to the functional properties of HEOs. STEM bridges this gap with its atomic-scale resolution and energy-selective imaging capabilities. [263–265]

STEM to directly observe nanoscale structural features in $(MgCoNiCuZn)_{0.2}O$ thin films, including cation-rich domains and the local emergence of distinct symmetry configurations.[41,268] Additionally, they demonstrate that HAADF-STEM can be used to generate strain maps, which are essential for understanding mechanisms of ordering evolution. These strain maps can then be integrated into developing complementary phase-field models, significantly enhancing predictive capabilities

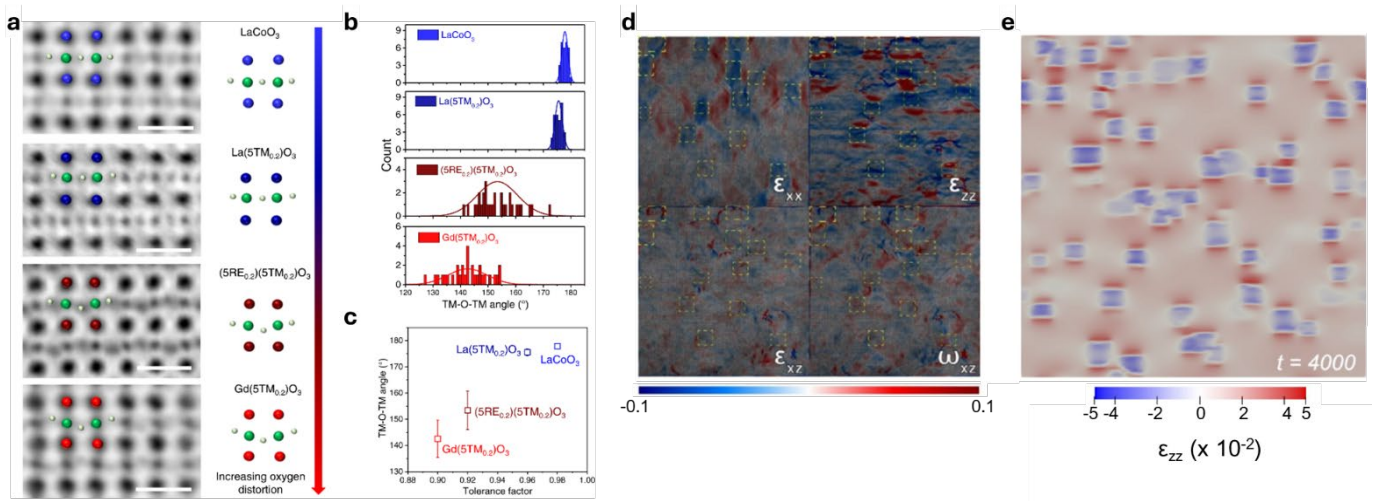

Figure 7: a) Annular bright field images of $LaCoO_3$, $La(CrMnFeCoNi)_{0.2}O_3$, $(LaNdSmGdY)_{0.2}(CrMnFeCoNi)_{0.2}O_3$, and $Gd(CrMnFeCoNi)_{0.2}O_3$, respectively, showing the transition metal–oxygen–transition metal (TM-O-TM) angle in their lattice structures. The green and silver spheres representing the B-site atoms and oxygen atoms, respectively, while the blue, dark blue, dark red and red spheres representing the A-site atoms in the four samples, respectively. Scale bar is 0.5 nm. b) Distribution of the TM–O–TM angle in the four samples. c) The relationship between TM–O–TM angle and tolerance factor. Data are presented as means of ± standard deviation of the distribution of TM–O–TM angle. d) Strain mapping with geometric phase analysis (GPA) on atomically resolved STEM images for spinel nano-cuboids (yellow boxes), e) phase field out-of-plane strain maps for spinel nano cuboids after 4000-time steps. Figures a through c were reproduced though a CC-BY license from Su, et al. Nat Commun 13, 2358 (2022). Figures d and e were adapted though a CC-BY from Almishal, et al. J Am Ceram Soc. 108, e20223 (2025).

*4.2.1 Scanning Transmission Electron Microscopy* STEM employs a focused electron beam scanned across the sample, providing high spatial resolution. Its primary imaging modes—bright-field (BF), annular dark field (ADF), and high-angle annular dark field (HAADF)—offer distinct contrast mechanisms. HAADF-STEM, utilizing Rutherford scattering sensitive to atomic number (Z-contrast), effectively visualizes heavier metal cations but poses challenges for lighter elements like oxygen. Despite this limitation, HAADF-STEM has significantly contributed to understanding HEOs structures and morphologies. In Figures 7a-c, Su *et al*. revealed local ordering in high entropy perovskites using ABF to elucidate relationships between TM-O bond angles.[266] Huang *et al*. identified preferential ordering along specific crystallographic planes in $(CeYLaHfTiZr)_xO_\delta$ system.[267] Furthermore, Huang *et al*. used ADF to examine increased oxygen vacancies near HEO nanoparticle surfaces.[267] Almishal *et al*. utilized HAADF-and accurately reproducing the experimentally observed nanoscale ordering. Figures 7d,e show an example of such HAADF-STEM strain maps for spinel-nano cuboids in the $(MgCoNiCuZn)_{0.2}O$ thin films and the corresponding simulated strain maps from phase field modeling. Several studies have employed HAADF-STEM in combination with EELS to investigate elemental distribution and oxidation states in HEOs. For example, the aforementioned spinel nano cuboids were found to exhibit elevated concentrations of $Co^{3+}$ [41], while in $CuCeO_x$–HEO nanoparticle systems, $CeO_2$ particles were observed to graft onto the surface of HEO particles[269]. These techniques have also been used to probe strain and cation distribution at grain boundaries, often associated with oxygen deficiencies.[270]

Emerging STEM-based techniques like integrated differential phase contrast (iDPC-STEM), annular bright-field (ABF-STEM), and four-dimensional STEM (4D-STEM) offer complementary analytical strengths. ABF-STEM significantly improves oxygen detection and facilitates detailed oxygen vacancy mapping. Additionally, machine learning methods have become crucial for analyzing extensive and complex TEM datasets, providing rapid and quantitative identification of subtle structural and chemical ordering. Therefore, the integration of electron microscopy, advanced analytical techniques, computational modeling, and machine learning establishes a comprehensive framework for resolving the intricate structural and chemical complexities inherent in HEOs at the local scale. Such multi-scale, multi-modal approaches are essential for engineering next-generation oxide materials with tailored functionalities.

*4.2.2 Atom Probe Tomography.* APT is a hybrid microscopy technique that provides a compositional map with three-dimensional spatial resolution and chemical sensitivity. Samples are fabricated in a needle-like structure with an apex tip radius typically < 100 nm and held at cryogenic temperatures in an ultra-high vacuum chamber to minimize unprompted evaporation and ionization. Measurements are facilitated by high-voltage pulses in conjunction with femtosecond laser pulses to systematically evaporate ions from the sample. Once evaporated, ions travel up to 200 mm to a position-sensitive detector comprised of microchannel plates to elucidate initial position and time-of-flight (TOF). The TOF data correlated to the mass-to-charge ratio of ions to resolve the chemical species of each impact through the relation $m/n = 2eVt^2/L^2$, which allows for this to be applied to the full range of the periodic table. The detector records the ($x,y$) position of the impacts and reverse projects the positions to a dynamically updated quasi-stereographic projection model to determine the $z$-position of the ions. This technique provides truly atomic-scale precision as the spatial resolution is 0.2-0.3 nm laterally and 0.1-0.2 nm in depth.[271]

Studies have demonstrated the power of APT in elucidating nanoscale chemical fluctuations and phase formation mechanisms in HEO systems. Correlative APT and TEM to reveal that controlled cooling of (MgCoNiCuZn)$_{0.2}$O leads to partial phase separation and local clustering, particularly of Cu, even when XRD indicates a nominally single-phase structure.[214] These subtle chemical redistributions underscore the need for atomic-scale compositional analysis in interpreting stability and functional behavior. Building on these insights, Dupuy et al. combined APT and TEM to examine the nucleation and growth of secondary phases in (MgCoNiCuZn)$_{0.2}$O.[212] They observed that Cu-rich and Co-rich regions, initially maintaining rocksalt-like structure, served as precursors to the formation of crystalline tenorite and spinel , respectively. These transformations were spatially localized and facilitated by cation vacancies and local structural distortion. Importantly, elemental segregation preceded crystallization, supporting a non-classical nucleation pathway, and growth was governed by cation diffusion through the parent matrix. Gwalani *et al.* applied APT to a $Y_2O_3$-doped high-entropy alloy and discovered the in-situ formation of compositionally complex nanoscale oxide dispersoids.[188] The oxides exhibited high configurational complexity with contributions from multiple cation species, supporting the notion of entropy-stabilized secondary phases. Collectively, these results reinforce APT's unique ability to detect early-stage, nanoscale inhomogeneities that influence macroscopic behavior in HEOs.

## 4.3 Total Scattering

Total scattering is a technique that collects the entire scattering signal from a material to resolve the structure on multiple length scales. The intricacies of total scattering are outside the scope of this review, but in-depth discussion of them can be found in books and reviews.[272,273] To be brief, the scattering vector, Q, is combined with atomic specific variables to create the scattering function S(Q). It can be seen in $Q = 4\pi sin(\theta)/\lambda$ how sharp peaks will appear when the Bragg condition for constructive interference, $1/d = 2sin(\theta)/n\lambda$, is fulfilled.[274] The periodic lattice positions are captured in this Bragg diffraction to give the long-range order of the system. The deviations from the perfect lattice though local disorder, defects, and short-range order are communicated in the diffuse scattering. S(Q) captures both signals in reciprocal space and is Fourier Transformed to create the pair distribution function (PDF) of interatomic distances in real space. Being a Fourier Transform, the maximum resolution of the PDF will be achieved by maximizing the range of Q that is probed.

*4.3.1 X-ray Scattering.* Using X-rays as the scattering radiation to collect PDF (xPDF) data means that the atomic form factor must be taken into account. X-rays interact with the electronic orbitals around nuclei and the atomic form factor is a function of the electron density as well as Q, meaning it will scale with Z and change with the scattering angle. This form factor limits the momentum transfer space available in xPDF due to the fact that the intensity of scattering decreases as angle increases.[274] The high flux of photons from synchrotron sources can make measurement feasible at higher angles and have the benefit of shorter collection times.[272,274] Laboratory xPDF instruments are also available that utilize sources with shorter wavelengths, such as Mo or Ag, to increase the Q range.[272]

*4.3.2 Neutron Scattering.* Neutrons scatter off of the nuclei of atoms, giving neutron PDF (nPDF) measurements the leg up on xPDF in a couple of aspects. The smaller size of the nuclei compared to the electron orbitals renders the atomic form factor to 1 for all Q, maximizing the momentum transfer space that is visible.[274] Nuclei also have characteristic cross-sections and scattering lengths that allow for sensitivity to light elements, down to H.[275] Neutrons have spin 1/2, giving them a magnetic moment and allowing them to probe the magnetic structure of materials based on the magnetic interaction operator and the magnetic form factor.[276] These aspects are especially important for HEOs where there is a mixing of cations with varying levels of magnetism, the common inclusion of Mg or Li, and dislocations under an angstrom.

*4.3.3 PDF Analysis.* The local disorder characteristic to HEOs makes total scattering a powerful technique for elucidating the structure of the system. Least-squares refinement of the real space data can be used to fit structural parameters of a model, including the unit cell, atomic displacements, and occupations. PDFfit2 is an application to do this built on Python and has the user friendly PDFgui built upon it that includes macros for temperature, doping, and r series fitting.[277] Another analysis technique is using Reverse Monte Carlo simulations to refine structural models based on the total scattering data. An example of such simulations combined with neutron experimental data, G(r), and partial pair correlations is shown in Figure 8. RMCProfile is a program that can do this for a supercell, building the model using individual atomic moves and outputting the structure along with information such as atomic displacements, and has recently been updated to allow the simultaneous consideration of both total scattering and EXAFS data.[278] The power of total scattering to discover differences in the local structure of HEOs from the average structure has been utilized in numerous studies to date. A study by Jiang et al. will be discussed in more detail later, but there have been multiple studies where local distortions are found in high-symmetry structures.[62,279]

### 4.4 Computational Methods

Computational methods are essential in helping understand the local structure of novel materials through system modeling. There are a variety of approaches to predict the local structure of a material using principles of quantum mechanics and statistical mechanics to model atomic interactions. Methods such as molecular dynamics, density functional theory, and Monte Carlo simulations have been used to study high entropy materials.

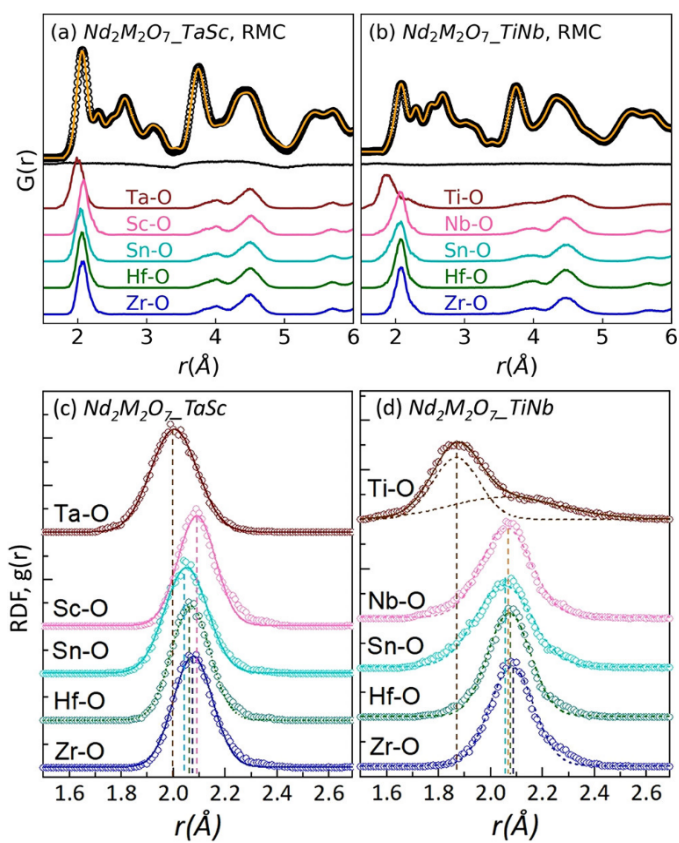

Figure 8: Experimental G(r) (black circles) and Reverse Monte Carlo simulation (orange lines) for (a) $Nd_2M_2O_7$ with M = Ta,Sc,Sn,Hf,Zr (Nd2M2O7_TaSc), and (b) $Nd_2M_2O_7$ with M = Ti, Nb, Sn, Hf, Zr ($Nd_2M_2O_7$_TiNb), with partial pair correlation functions from models shown below as separate, labeled, colored lines. Gaussian peak fitting of the first M–O pair–pair correlation peaks from the RMC partial PDF distributions in (c) Nd2M2O7_TaSc and (d) $Nd_2M_2O_7$_TiNb. A double deconvoluted Gaussian peak was used to fit the Ti–O bond distribution peaks in $Nd_2M_2O_7$_TiNb. Reprinted with permission from Jiang et al. J. Am. Chem. Soc. 2021, 143, 11, 4193-4204. Copyright 2021, ACS.

Molecular dynamics (MD) is a computational model that solves Newton's equations of motion to equilibrate the system based on approximate effective potentials.[280] This method has been used to model edge dislocation density in HEOs[281] and study oxygen self-diffusion in a fluorite-structured HEO.[282] This method has also been used to predict properties of a high entropy carbide with good correlation to experimental results.[283] Due to the complexity of interactions present in high entropy materials, *ab initio* molecular dynamics (AIMD) may be an advantageous route to follow. Where MD uses Newton's equations, AIMD uses the Schrodinger equation to allow for real potentials to model quantum effects.[280] AIMD has been used to model thermal fluctuations that inform density functional theory calculations to simulate local environments in HEOs.[284]

Density functional theory (DFT) is a method of computing the electronic structure of materials using electron density rather than wave function theory, which reduces the computational cost of modeling systems.[285] DFT can be used to predict thermodynamic stabilities, electronic structures, and defect formation energies with good accuracy.[286] Due to the large configurational space of HEOs it is difficult to test an entire range of material compositions, but DFT can help screen desirable candidates.[287] DFT has been used in this manner to help investigate a novel catalyst composition for an oxygen evolution reaction.[288] Vienna *ab initio* simulation package (VASP) is a DFT software which uses plane-wave approximations and is commonly applied to HEOs.[289–291]

Monte Carlo simulations are a statistical technique used to predict outcomes by modeling random processes and estimating probabilities based on input parameters.[292] This method generates a range of possible results by varying input factors, such as atomic configurations, energy states, or temperatures, according to predefined probability distributions.[292] These simulations are useful for modeling HEO systems, where factors like disorder and multi-component mixing play crucial roles. Monte Carlo methods have been applied to predict material behaviors such as magnetic response[293] and component segregation as a function of temperature.[291] A Monte Carlo model developed for HEO perovskites by Motseyko *et al.* showed that the order-disorder transition temperature ($T_{od}$) correlates with the standard deviation of cation-anion bond lengths, offering a simple way to estimate $T_{od}$ in multi-cationic materials like HEOs.[294]

## 5. HEO Characterization in Literature: Select Case Studies

When investigating the novel phenomena that occur in HEOs, the observed property changes often admit competing explanations. With the disorder in these materials spanning a broad spectrum of modes and length scales, one measurement can often miss or misattribute the origin of a feature. Only when corroborating multiple data sets do the pieces of the puzzle fit together. An excellent compilation of work by Sarkar *et al.* demonstrates the complexity of experimentally characterizing an HEO system at all length scales to enable a full-scale picture of the material and its subsequent properties.[220] Figure 9 depicts their experimental strategy. Here, we present examples from literature where the application of complementary characterization methods reveals how high entropy effects manifest themselves in unpredictable, but insightful, ways.

Jiang *et al.* investigated a series of Nd-based pyrochlore HEOs ($Nd_2M_2O_7$) that were synthesized to understand how the cation disorder and local distortions affect the structure-property relationships.[284] The techniques used include nPDF, reverse Montecarlo simulations, and DFT calculations. Neutron scattering and PDF analysis revealed that both compositions crystallized in a lower symmetry orthorhombic pyrochlore (*Imma*) structure. These techniques demonstrated that while $Nd_2(Ta_{0.2}Sc_{0.2}Sn_{0.2}Hf_{0.2}Zr_{0.2})_2O_7$ exhibited a relatively homogeneous distribution of cations, $Nd_2(Ti_{0.2}Nb_{0.2}Sn_{0.2}Hf_{0.2}Zr_{0.2})_2O_7$ showed significant $TiO_6$ octahedral distortions and chemical short-range order (SRO). The distortions in the latter composition were attributed to clustering effects involving Ti and the associated J-T distortions. RMC simulations further supported these findings by capturing statistical distributions of cation positions, highlighting distinct local chemical environments for the respective M-site cations. Additionally, DFT and AIMD simulations identified variations in local bond lengths and octahedral distortions, with specific trends in M-O bond distributions correlating with cation characteristics. For example, Ti cations displayed pronounced local distortions compared to other cations, which aligned with PDF and RMC results. Tolerance factor analysis showed that subtle differences in cationic radii and bonding preferences between the two compositions contributed to variations in their structural behavior. Specifically, Ta-Sc pairs in one compound exhibited stronger local associations, while Ti clustering in the other compound dominated the observed short-range order. These findings highlight the critical role of cation selection in tuning the local structure of HEOs, offering a framework to engineer specific properties such as ionic conductivity, magnetism, or catalytic activity.

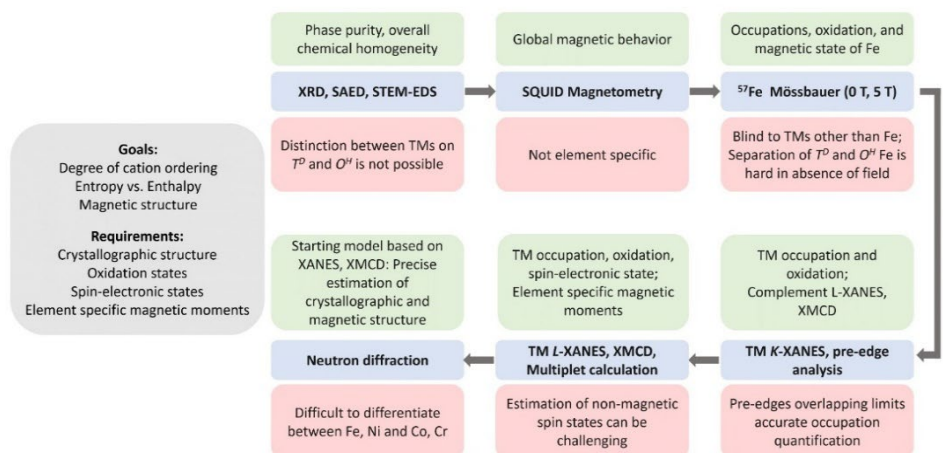

Figure 9: A schematic representation of the comprehensive experimental protocol used to achieve the picture of enthalpy/entropy balance and magnetic structure in spinel-based HEO ($Co_{0.2}Cr_{0.2}Fe_{0.2}Mn_{0.2}Ni_{0.2})_3O_4$. The limitations of the individual techniques to provide a complete solution in the case of HEOs are indicated in the light pink boxes, which indicate the necessity of the combined and cross-correlated experimental techniques used here. The conclusions that are drawn from each characterization set are highlighted in the light green boxes. Reproduced from Sarkar *et al.* Acta Materialia, 226, 2022, 117581 with permission from Elsevier.

Min *et al.* utilized a wide range of characterization techniques to discern the local magnetic ordering and local environment across the composition space of A-site disordered ferrite spinel single crystals.[295] Samples were created according to the formula $A_xFe_{3-x}O_4$, where A represents the equimolar mixture of Mg, Mn, Co, Ni, and Fe and $x = 1, 1.5, 1.8$. The S1, spinel of x=1, and S1.5 samples retained a high crystal quality in XRD measurements of powdered crystals while the S1.8 sample displayed a splitting of peaks associated with lattice distortions; and energy-dispersive X-ray spectroscopy (EDS) showed that there was global chemical homogeneity. Previous studies show that single-element substitutions in spinels suppress magnetic ordering and broaden magnetic transitions. The large degree of lattice and local structure distortion inherent to high entropy oxides is expected to create broad transitions as a function of temperature and gradual saturations in magnetic hysteresis. The high configurational entropy within these samples manifests in surprising magnetic characterization results. Field-cooling measurements under 100 and 1000 Oe show that the S1 and S1.5 transitions are comparably slim to the $Fe_3O_4$ standard. There is a trend of depressed transition temperature with an increased ratio of disorder and S1.8 having the broadest transition corresponding to the lower degree of crystallinity. The magnetic hysteresis displays slightly decreased saturation $\mu_B$ values with increased chemical disorder at 300 K, and nearly identical at 5 K. The room temperature measurement shows that saturation is reached by 0.5 T and steeper saturation curves than the standard at 5 K for the S1 and S1.5 samples. Again, the structural disorder of S1.8 creates gradual saturation, but it reaches a similar value to S1.5. The suppression of $\mu_B$ compared to $Fe_3O_4$ suggests that there is antiparallel ordering of the tetrahedral and octahedral sites. Ferrimagnetic ordering is confirmed by the relative intensities of magnetic peaks in the neutron diffraction pattern. XAFS was employed to investigate the local structures of the samples to gain understanding of the trends seen in magnetism. When compared to $CoFe_2O_4$ and $NiFe_2O_4$, the Co and Ni absorbers in these samples mirror the trends of the inverted spinel standards with tetrahedral occupation slightly increasing with disorder, evidenced in the EXAFS. Fe remains in a mixed occupation state across the sample series while showing a higher ratio of $3^+/2^+$ in the XANES than pure $Fe_3O_4$. The Mn absorber shows evidence of a consistently mixed valence state in the XANES, trending towards higher values as the Fe fraction is decreased. In the EXAFS, Mn shifts from a majority tetrahedral occupation to octahedral in S1 to S1.8. STEM/EELS measurements corroborate the XANES findings of mixed valency in the Mn, Fe, and Co ions and high-resolution EDS reinforces the lack of short-range ordering that is seen in the single-element doped studies.

Zhang et al. studied the A-site disorder of a series of high entropy perovskites with the composition $(CaSrBaLaM)_{0.2}TiO_3$, where M = Nd, Sm, Dy, Y, or Pb, for thermoelectric applications with emphasis on the impact of local disorder on thermal and electrical transport properties.[296] While tolerance factor calculations predict cubic phase stability across the sample series, a secondary pyrochlore phase was persistent in the Nd, Sm, Dy, and Y containing samples according to XRD. Therefore, the single phase Pb containing composition was chosen for further characterization. Along with XRD, structural investigation was carried out with SEM-EDS, HAADF-TEM with EDS, and HRTEM. SEM imaging revealed uniform grains on the order of one micron at the determined optimal sintering temperature of 1250°C. SEM-EDS maps displayed uniform distribution of the A-site cations in equimolar proportions. HAADF with EDS and HRTEM showed that crystallinity and chemical distribution were retained when samples were post-annealed at 1300°C for electrical characterization. HRTEM revealed nanoscale grains (4-6 nm) with irregular orientations and edge dislocations, particularly at grain boundaries, introducing local strain fields. The combination of structural distortions, cation disorder, and strain fields significantly disrupted phonon transport, leading to a reduced lattice thermal conductivity of 1.75 W/m·K at 1073 K. These findings demonstrate how short-range disorder, induced by the high entropy design, enhances phonon scattering and reduces thermal conductivity, making the synthesized ceramics a promising candidate for thermoelectric applications.

Johnstone and González-Rivas et al. investigated the evolution of magnetic properties in high-entropy spinels with systematic substitution of non-magnetic Ga in the composition $(CrMnFeCoNi)_{3-x}Ga_xO_4$ with x=0, 0.3, 0.6, 1.2.[297] Using XMCD and powder neutron diffraction, they correlated structural changes with trends in magnetization and magnetic susceptibility. The x=0 sample ferrimagnetically orders at 403 K with a secondary transition occurring at 55 K attributed to magnetic depinning. Increasing Ga content suppresses the primary transition while increasing the secondary one, with the 40% Ga sample ordering at 121 K. The saturation magnetization, $\mu_{sat}$, of the pure sample is lower than predicted due to antiparallel A-B site alignment. $\mu_{sat}$ increases with 10-20% Ga before decreasing below the original value at 40%. Magnetic hardening is also observed with increasing Ga content, with a coercive field maximizing at 880 mT at 40% Ga.

XMCD measurements reveal site and valence-specific behaviors: $Cr^{3+}$ and $Ni^{2+}$ remain exclusively octahedral across the series; $Fe^{3+}$ shifts from primarily tetrahedral to octahedral; $Co^{2+}$ transitions from fully tetrahedral to Jahn-Teller-distorted octahedral. This change in Co coordination drives the observed magnetic hardening.[297] Mn acts as the charge compensator, shifting toward $Mn^{2+}$ as Ga content increases, accompanied by a transition from distorted octahedral to tetrahedral sites. In the 0% sample, Mn exhibits an average valence of ~$3.5^+$, decreasing to ~$2.5^+$ at 40% Ga. The XMCD-derived A- and B-site occupancies were used to estimate sublattice configurational entropy. While the total entropy exceeds the 1.5R high-entropy threshold, the octahedral site alone reaches this threshold with Ga substitution, and the tetrahedral site enters the medium-entropy regime (> 1R), peaking at x=0.6. These findings suggest that sublattice-specific entropy can be selectively tuned, with implications for tailoring material properties through targeted compositional design.

## 6. Future Directions

Local structure plays a defining role in determining the functional properties of HEOs, and continued progress in the field will rely on a deeper, more quantitative understanding of how chemical disorder manifests at the atomic scale. While a wider range of tools including XAFS, STEM/EELS, total scattering, and Raman spectroscopy among others, have proven invaluable, the true opportunity lies in integrating these techniques into unified, multi-scale characterization frameworks. The integration is essential to link subtle variations in local coordination, bond length distributions, and valence fluctuations to emergent properties such as ionic transport, magnetism, and electronic response.

A major future direction involves decoupling the contributions of disorder of different sublattices. In complex oxide systems such as spinels, perovskites, and pyrochlores, the A and B sites often host cations with different valences, ionic radii, and bonding preferences.

As demonstrated in the spinel and perovskite HEOs [295–297], the ability to selectively control disorder on one sublattice while maintaining order on another provides a unique mechanism to tailor properties. Future studies should explore the entropy balance between sublattices, using a combination of synchrotron-based EXAFS[60,145,298], neutron scattering[284,299], and configurational entropy calculations derived from site occupancy analysis via STEM-EDS/EELS[153,256,257,297]. Site-specific entropy tuning wherein only one sublattice meets the 1.5R threshold may enable property optimization without compromising structural coherence.

Understanding and controlling local structure is not only essential for decoding emergent behavior in high entropy oxides, but also has far-reaching implications across materials chemistry, condensed matter physics, and applied engineering. Concepts such as structural disorder and tunability are increasingly relevant to heterogeneous catalysis, neuromorphic computing, solid-state electrolytes, and radiation-tolerant ceramics. By bridging fundamental mechanisms with functional outcomes, local structure analysis serves as a foundation for materials innovation across disciplines.

Another promising area lies in advancing total scattering and PDF techniques. The capacity of total scattering to resolve structural information across both short- and intermediate-range order is particularly valuable in HEOs, where distortions often occur without changing average symmetry. PDF analyses, when performed using boxcar or variable-r fitting schemes, can reveal how the local environment evolves as a function of composition, thermal history, or external field. Coupling these with RMC simulations allows for atomistic reconstructions that reflect real-space disorder.[284] The simultaneous refinement of EXAFS and PDF data within the same structural model, now feasible using tools such as RMCProfile[278,300], represents a critical next step. Such joint modeling can enhance confidence in assigning coordination environments and quantifying bond length distributions.

In parallel, the role of the oxygen sublattice remains comparatively underexplored. Given the prevalence of charge compensation via oxygen vacancy formation in aliovalent-doped HEOs[25,151], future work should incorporate O $K$-edge XAS, neutron total scattering and Raman spectroscopy to systematically map oxygen coordination and vacancy clustering. These effects not only impact ionic conductivity and redox behavior[72], but may also influence magnetic ordering and dielectric response via lattice distortions. Advanced techniques such as aberration-corrected STEM coupled with 4D-STEM and ABF imaging may offer additional sensitivity to light elements like oxygen[267], helping to map subtle off-stoichiometry and detect soft-mode distortions.

A further direction involves the development of *in situ* and *operando* techniques to capture dynamic structural evolution under functional conditions. While most studies of local disorder rely on static measurements, real-world applications often involve high temperatures, electric fields, or reactive environments. The use of *in situ* XANES, Raman, or diffraction under applied stimuli (e.g., electric field, temperature, partial pressure) could uncover how local environments reconfigure during device operation or catalysis[73]. For example, tracking cation oxidation states and coordination environments during electrochemical cycling could clarify charge/discharge mechanisms and structural degradation pathways in energy storage applications.[96,152]

Advances in data science and machine learning (ML) offer transformative potential for accelerating the characterization of high entropy oxides. As the number of compositional permutations and structural configurations in HEOs grows exponentially, ML techniques can help uncover correlations between atomic-scale disorder and macroscopic properties. Recent studies have applied unsupervised clustering of XANES spectra, automated segmentation of STEM images, and high-throughput DFT screening to inform compositional design and phase stability trends[235]. However, integrating data across techniques remains a challenge, as each method is grounded in distinct physical principles and mathematical frameworks. Future progress will require the development of unified ML/AI platforms capable of simultaneously analyzing spectroscopy, diffraction, and microscopy data to quantify local disorder and predict functional performance. Such tools would not only accelerate interpretation but could also guide experimental design by identifying promising compositional regions with targeted specifically by disorder characteristics. A platform that seamlessly connects multi-modal datasets to structural models and property predictions would represent a major milestone in the field, one that could shift the paradigm from characterization to true materials-by-design.

## Conclusions

HEOs represent a rapidly growing frontier in materials science, where chemical disorder and configurational complexity give rise to remarkable functional behavior. Over the past decade, significant advances have been made in understanding how structural disorder, charge compensation, and lattice distortions contribute to the macroscopic properties of HEOs. However, much of this insight has focused on average structure or compositional trends. As this review has shown, the local structure encompassing short-range order, site-specific coordination, and nanoscale symmetry breaking is increasingly recognized as the true origin of many functional phenomena in HEOs.

In this context, the next generation of HEO research will hinge on our ability to bridge local structural motifs with property outcomes across thermal, electrical, magnetic, catalytic, and electrochemical domains. This effort will demand more than the individual application of advanced characterization tools; it will require synergistic integration of experimental, computational, and data-driven methodologies. Multi-modal investigations combining, for example, X-ray absorption spectroscopy, total scattering, STEM-EELS mapping, and DFT-based modeling, can offer a more complete picture of disorder, but only when interpreted through a unified framework.

Emerging tools in machine learning and artificial intelligence will be essential to meet this challenge. These methods offer the ability to correlate diverse datasets and identify patterns that may be invisible to conventional analysis. Future platforms capable of ingesting data from spectroscopy, diffraction, and microscopy simultaneously, while mapping structural features to functional behavior, could revolutionize how we design and evaluate new compositions. In this paradigm, local structure is not merely a static descriptor, it becomes a navigable design space.

Harnessing this potential will require a new level of collaboration across synthesis, theory, and characterization. As our understanding of disorder deepens, so too will our ability to exert predictive control over functionality, ushering in a new era of materials-by-design built not around idealized order, but around the creative engineering of disorder itself.

## Author contributions

John P. Barber: conceptualization, data curation, writing-original draft, writing-review & editing, supervision. William J. Deary: data curation, writing-original draft. Andrew N. Titus: data curation, writing-original draft. Gerald R. Bejger: data curation, writing-original draft. Saeed A.I. Almishal: data curation, writing-original draft, writing-review & editing. Christina M. Rost: conceptualization, data curation, visualization, writing-original draft, writing-review & editing, funding acquisition, supervision.

## Conflicts of interest

## Data availability

## Acknowledgements

The authors gratefully acknowledge the support from NSF through the Materials Research Science and Engineering Center DMR-2011839. J.P.B, G.R.B, and C.M.R also acknowledge partial support from NSF through the Faculty Early Career Development Program (CAREER), DMR-2440979.